\documentclass[12pt]{article}
\usepackage{amssymb}
\usepackage{graphicx}

\renewcommand{\theequation}{\thesection.\arabic{equation}}

\newlength{\extraspace}
\newlength{\extraspaces}
\newcommand{\be}{\begin{equation}
\addtolength{\abovedisplayskip}{\extraspaces}
\addtolength{\belowdisplayskip}{\extraspaces}
\addtolength{\abovedisplayshortskip}{\extraspace}
\addtolength{\belowdisplayshortskip}{\extraspace}}
\newcommand{\ee}{\end{equation}}

\newcommand{\ba}{\begin{eqnarray}
\addtolength{\abovedisplayskip}{\extraspaces}
\addtolength{\belowdisplayskip}{\extraspaces}
\addtolength{\abovedisplayshortskip}{\extraspace}
\addtolength{\belowdisplayshortskip}{\extraspace}}
\newcommand{\ea}{\end{eqnarray}}

\newcommand{\newsection}[1]{
\vspace{12mm}
\pagebreak[3]
\addtocounter{section}{1}
\setcounter{equation}{0}
\setcounter{subsection}{0}
\setcounter{footnote}{0}
\noindent{\bf \thesection. #1}
\nopagebreak
\medskip
\nopagebreak}

\newcounter{saveeqn}
\newcommand{\alpheqn}{\setcounter{saveeqn}{\value{equation}}%
 \stepcounter{saveeqn}\setcounter{equation}{0}%
 \renewcommand{\theequation}
     {\thesection.\mbox{\arabic{saveeqn}\alph{equation}}}}
\newcommand{\reseteqn}{\setcounter{equation}{\value{saveeqn}}%
  \renewcommand{\theequation}{\thesection.\arabic{equation}}}

\begin{document}
\addtolength{\baselineskip}{1.5mm}

\thispagestyle{empty}
\begin{flushright}
hep-th/0306044\\
\end{flushright}
\vbox{}
\vspace{2.5cm}

\begin{center}
{\LARGE{Multi-black hole solutions in five dimensions
        }}\\[16mm]
{H.~S. Tan~~and~~Edward Teo}
\\[6mm]
{\it Department of Physics,
National University of Singapore, 
Singapore 119260}\\[15mm]

\end{center}
\vspace{2cm}

\centerline{\bf Abstract}\bigskip
\noindent
Using a recently developed generalized Weyl formalism, we construct
an asymptotically flat, static vacuum Einstein solution that describes a
superposition of multiple five-dimensional Schwarzschild black holes.
The spacetime exhibits a $U(1)\times U(1)$ rotational symmetry.
It is argued that for certain choices of parameters, the black holes
are collinear and so may be regarded as a five-dimensional generalization
of the Israel-Khan solution. The black holes are kept in equilibrium by 
membrane-like conical singularities along the two rotational axes; however, 
they still distort one another by their mutual gravitational attraction. 
We also generalize this solution to one describing multiple charged 
black holes, with fixed mass-to-charge ratio, in Einstein-Maxwell-dilaton 
theory. 


\newpage

\newsection{Introduction}

The Israel-Khan solution describing multiple collinear Schwarzschild 
black holes in four dimensions has been known for some time \cite{Khan}. 
They belong to a class of static, axisymmetric solutions first obtained 
by Weyl \cite{Weyl} who showed that the corresponding vacuum Einstein
equations can be reduced to solving the Laplace equation in three-dimensional
flat space. Although the Israel-Khan solution contains string-like conical 
singularities, it nonetheless has a well-defined gravitational action 
\cite{Gibbons_Perry} and this enables one to study their interactions using 
standard techniques of gravitational thermodynamics \cite{Costa_Perry}.  

Recently, Emparan and Reall \cite{Emparan_Reall} generalized the Weyl 
formalism to arbitrary dimensions $D\geq 4$. They showed that the general 
solution of the $D$-dimensional vacuum Einstein equations which has 
$D-2$ orthogonal commuting isometries is specified by $D-3$ 
axisymmetric solutions of the Laplace equation in three-dimensional
flat space. A way to classify these solutions was also presented in
\cite{Emparan_Reall}. In particular, the five-dimensional (5D) Schwarzschild 
black hole, like its four-dimensional counterpart, belongs to the
generalized Weyl class. Another noteworthy member of this class is the
5D black ring solution \cite{Emparan_Reall}, which is the first 
example of an asymptotically flat vacuum spacetime with an event horizon
of non-spherical topology.

The generalized Weyl formalism opens up, for the first time, the 
possibility of obtaining multi-Schwarzschild black hole solutions in 
five dimensions. Three possible configurations were briefly discussed
in \cite{Emparan_Reall}. Firstly, a two-black hole solution was constructed,
with either black hole located at the north and south poles of a 
Kaluza-Klein bubble.\footnote{A detailed study of this solution
recently appeared in \cite{Elvang}.} This solution is not
asymptotically flat, since one coordinate is asymptotically a 
Kaluza-Klein circle. The second solution considered in \cite{Emparan_Reall}
was a three-black hole solution that is asymptotically flat. However,
it is not a collinear system, since the central black hole is only 
collinear with each of the other two black holes along {\it different\/} 
axes. Finally, a solution describing an infinite periodic array of black 
holes was also considered, although it was argued that this solution 
cannot be interpreted as a black hole localized on a Kaluza-Klein circle
because it does not have the correct asymptotic structure.

It would be very interesting to find a multi-black hole solution in five 
dimensions that may be considered a generalization of the four-dimensional
Israel-Khan solution. Such a solution should satisfy two conditions.
Firstly, it should be asymptotically flat, rather than asymptotically 
Kaluza-Klein as in the first solution mentioned above. Secondly, it should 
describe a `collinear' array of black holes, which rules out the second 
solution. Such a notion of collinearity would have to be compatible with the 
spatial symmetries imposed on 5D Weyl solutions, i.e., $U(1)\times U(1)$ 
corresponding to the two orthogonal commuting rotational Killing vectors.

In this paper, we construct a multi-Schwarzschild black hole solution 
(which is actually the finite version of the infinite black hole solution 
considered in \cite{Emparan_Reall}) which we argue, satisfies the
above two conditions and thus qualifies as a 5D analog of the
Israel-Khan solution. We begin in Sec.~2 with the explicit construction
of our solution and a characterization of the conditions under which it 
may be considered a collinear array of black holes. In Sec.~3, we 
examine some basic properties of our solution for the case of a 
two-black hole system. In particular, we perform various limiting 
procedures and study the near-horizon geometries of the black holes, 
which turn out to be distorted by their mutual gravitational attraction. 
In Sec.~4, the analysis is briefly repeated for the three-black hole 
system. In Sec.~5, the charged version of our solution in the framework 
of Einstein-Maxwell-dilaton theory is obtained and studied. The extremal
limit is then examined in Sec.~6. The paper ends with a discussion of
some possible extensions of this work.

\newsection{The multi-Schwarzschild black hole solution}

In this paper, we are specifically interested in 5D static spacetimes 
belonging to the generalized Weyl class, i.e., possessing three orthogonal 
commuting Killing vectors. Such a spacetime metric can be written as
\be
\label{metric}
{\rm d}s^2=-{\rm e}^{2 U_1}\,{\rm d}t^2+{\rm e}^{2 U_2}\,{\rm d}\varphi^2 + {\rm e}^{2 U_3} \, {\rm d}\psi^2 + {\rm e}^{2\nu} ({\rm d}r^2+{\rm d}z^2)\,,
\ee
where $\nu=\nu(r,z)$ and $U_\alpha=U_\alpha(r,z)$ for $\alpha=1,2,3$. 
The vacuum Einstein equations can be shown to reduce to the Laplace 
equation:
\be
\label{Laplace}
\frac{\partial^2 U_\alpha}{\partial r^2} 
+ \frac{1}{r}\frac{\partial U_\alpha}{\partial r} 
+ \frac{\partial^2 U_\alpha}{\partial z^2} = 0\,,
\ee
with $\nu$ determined by quadratures \cite{Emparan_Reall}. The three 
harmonic functions $U_\alpha$ can be thought of as the Newtonian potentials 
produced by rods of zero thickness and density $\frac{1}{2}$ along the 
$z$-axis.  They should add up to the potential of an infinite 
rod. An important example of a spacetime in this class is the 5D 
Schwarzschild solution, which has the potentials of the rod structure
as shown in Fig.~1.

\begin{figure}
\begin{center}
\includegraphics{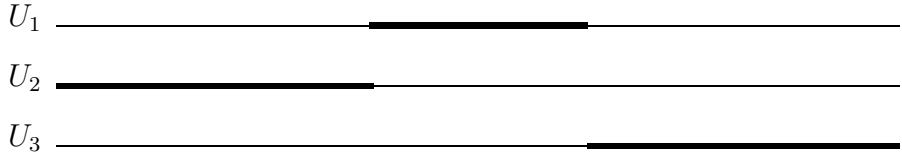}
\caption{Rod structure of the 5D Schwarzschild black hole solution.}
\end{center}
\end{figure}

As described in \cite{Emparan_Reall}, certain important properties of 
a generalized Weyl spacetime can be read off from its corresponding rod 
structure, even without the explicit form of the metric. For example, 
rod sources for the two angular coordinates, $\varphi$ and $\psi$, 
correspond to fixed points of these rotations, i.e., the symmetry axes. 
If the rod sources for the $\varphi$ and $\psi$ coordinates extend to 
infinity in either direction, then the spacetime is asymptotically flat. 
Another important fact is that a finite rod source for the time coordinate 
corresponds to an event horizon in the spacetime. 
Moreover, if either end of this rod continues with rods of
{\it different\/} angular coordinates, then the event horizon will 
have $S^{3}$ topology. This means that the rod structure in Fig.~1
describes a black hole in an asymptotically flat spacetime, in agreement
with the physical interpretation of the 5D Schwarzschild solution.

Bearing these facts in mind, let us now attempt to draw a rod structure 
corresponding to a superposition of $N$ Schwarzschild black holes that
generalizes the Israel-Khan solution. We first note that it must have
$N$ finite rods for the time coordinate, corresponding to $N$ disconnected
event horizons. Furthermore, to ensure that each horizon has $S^3$ topology,
the ends of each rod must continue with rods of different angular 
coordinates. Finally, the rod structure should have the same 
asymptotic form as that in Fig.~1, to ensure asymptotic flatness. 
These conditions leave us with the rod structure in Fig.~2 as one of 
the simplest possibilities. Of course, other rod structures are also 
possible; however, we do not consider them to be as natural or 
compelling as the one chosen above.

\begin{figure}
\begin{center}
\includegraphics{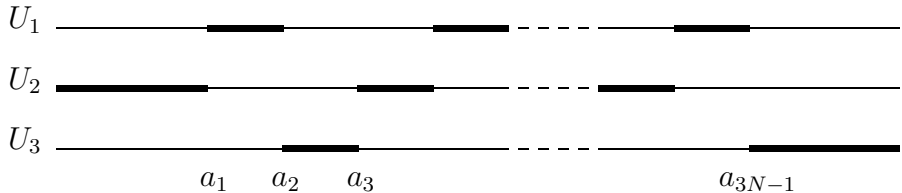}
\caption{Rod structure of an $N$-Schwarzschild black hole solution.}
\end{center}
\end{figure}

It is straightforward to write down the $U_\alpha$'s corresponding 
to the rod structure in Fig.~2. If we label the locations of the rod-ends 
in order of increasing $z$ (left to right in Fig.~2) by $a_1$, $a_2$, ..., 
$a_{3N-1}$, then we have
\alpheqn
\ba
\label{U_alpha}
&&U_1= \frac{1}{2} \sum_{k=1}^N \log \frac{R_{3k-2}-\zeta_{3k-2}}
{R_{3k-1}-\zeta_{3k-1}}\,,\\
&&U_2= \frac{1}{2} \sum_{k=2}^N \log \frac{R_{3k-3}-\zeta_{3k-3}}
{R_{3k-2}-\zeta_{3k-2}} + \frac{1}{2} \log(R_1+\zeta_1)\,,\\
&&U_3= \frac{1}{2} \sum_{k=2}^N \log \frac{R_{3k-4}-\zeta_{3k-4}}
{R_{3k-3}-\zeta_{3k-3}} + \frac{1}{2} \log(R_{3N-1}-\zeta_{3N-1})\,,
\ea
\reseteqn
where $R_i \equiv \sqrt{r^2 + \zeta_i^2}$ and $\zeta_i \equiv  z - a_i$, 
for  $1\le i \le 3N-1$. Using the method described in \cite{Emparan_Reall}, 
we can then solve for $\nu$. After some calculation, we obtain the line 
element for this rod structure to be:
\ba
\label{general metric}
{\rm d}s^2&=&-\prod_{k=1}^{N} \biggl( \frac{R_{3k-2}-\zeta_{3k-2}}
{R_{3k-1}-\zeta_{3k-1}}\biggr)\,{\rm d}t^2
+(R_1+ \zeta_1)\prod_{k=2}^{N}\biggl(\frac{R_{3k-3}-\zeta_{3k-3}}
{R_{3k-2}-\zeta_{3k-2}}\biggr)\,{\rm d}\varphi^2  
\cr
&&+(R_{3N-1}-\zeta_{3N-1})\prod_{k=2}^{N} \biggl(\frac{R_{3k-4}-\zeta_{3k-4}}
{R_{3k-3}-\zeta_{3k-3}}\biggr)\,{\rm d}\psi^2 
+  {\rm e}^{2\gamma_0} \frac{ \sqrt{Y_{1,3N-1}(R_{3N-1}-\zeta_{3N-1})}}
{R_1R_2\cdots R_{3N-1} \sqrt{R_1-\zeta_1}}
\cr
&&\times \prod_{k=2}^{N} \frac{\sqrt{Y_{3k-2,3N-1}Y_{3k-3,3N-1}Y_{1,3k-3}
Y_{1,3k-4}Y_{3k-3,3k-2}Y_{3k-4,3k-3}Y_{3k-4,3k-2}}}{Y_{3k-4,3N-1}Y_{1,3k-2}}  
\cr
&&\times \prod_{1<k<j} \frac{\sqrt{Y_{3k-3,3j-2}Y_{3k-2,3j-3}Y_{3k-4,3j-3}
Y_{3k-3,3j-4}Y_{3k-2,3j-4}Y_{3k-4,3j-2}}}{Y_{3k-2,3j-2}Y_{3k-3,3j-3}Y_{3k-4,3j-4}} 
\cr
&&\times \prod_{k=2}^N \frac{\sqrt{R_{3k-4}-\zeta_{3k-4}}}
{\sqrt{R_{3k-2}-\zeta_{3k-2}}}\,({\rm d}r^2+{\rm d}z^2)\,,
\ea
where $Y_{ij} \equiv R_iR_j+\zeta_i\zeta_j+r^2$, and ${\rm e}^{2\gamma_0}$ 
is a constant to be adjusted appropriately below. Note that this 
solution contains $3N-2$ free parameters, with $N$ parameters related to 
the individual masses of the black holes and the rest determining their 
spatial arrangement. 

Before embarking on a study of their spatial arrangement, let us consider 
the background spacetime limit of (\ref{general metric}) in which all the 
black holes disappear. This corresponds to shrinking the rods for the time 
coordinate down to zero size, leaving just the rods for the $\varphi$ and 
$\psi$ coordinates as in Fig.~3. It can be seen that the resulting rod 
structure corresponds to the Euclidean version of the multiple C-metric 
solution derived in \cite{Dowker} (in this case describing $N-1$ 
accelerating black holes), with a flat time direction added on. This 
spacetime is clearly non-flat when $N\geq2$. In addition to the usual two
semi-infinite rotational axes for the $\varphi$ and $\psi$ coordinates,
it contains $N-1$ finite-length rotational axes for each coordinate.
Observe that these axes are not one-dimensional lines, but rather
two-dimensional membranes. While the semi-infinite axes have the topology 
of open disks $D^2$, the finite ones have the topology of spheres $S^2$. 
Furthermore, from the behavior of the multiple C-metric, we know that there 
are in general conical singularities running along the axes. If we demand 
the two semi-infinite axes to be regular, then there are unavoidable conical 
singularities along the finite axes. Thus, this spacetime consists of 
$2(N-1)$ conical membranes with $S^2$ topology. They are orthogonal to
one another, in the sense that any line of constant longitude of one $S^2$
is orthogonal to any line of constant longitude of an adjacent $S^2$ 
at their adjoining point.

\begin{figure}
\begin{center}
\includegraphics{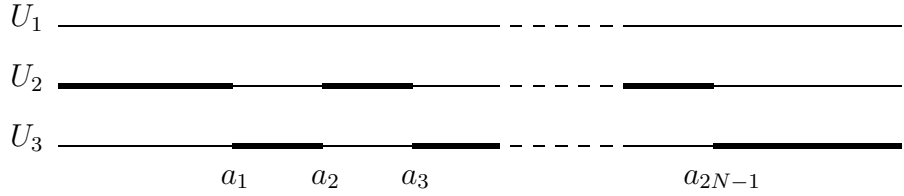}
\caption{Rod structure of the background spacetime of (\ref{general metric}).}
\end{center}
\end{figure}

At first, it may seem rather strange to have such a non-trivial spacetime
as the background. However, this is basically forced upon us if we want to 
construct multiple black hole solutions within the generalized Weyl 
formalism, and can be seen as follows. A black hole with an event horizon
of $S^3$ topology can only be introduced at points along the $z$-axis 
where the rods for the $\varphi$ and $\psi$ coordinates meet, i.e., 
fixed points of the $U(1)\times U(1)$ rotational symmetry. There can 
only be one such point in 5D Minkowski space \cite{Emparan_Reall}. If 
we require two or more fixed points, then the only possible asymptotically 
flat background, with a flat time direction, is the Euclidean multiple 
C-metric solution as described above. A solution corresponding to $N-1$ 
accelerating black holes has $2N-1$ such fixed points, although only $N$ 
of them were used in the construction of the solution 
(\ref{general metric}). These points are labeled $a_1$, $a_3$, ..., 
$a_{2N-1}$ in Fig.~3.

The reason for choosing these $N$ points alternately, is because any
three adjacent fixed points cannot be collinear. For example, $a_1$ and $a_2$
are collinear along the $\psi$-axis (with an $S^2$ conical membrane 
connecting them), while $a_2$ and $a_3$ are collinear along 
the $\varphi$-axis (with another $S^2$ conical membrane connecting them). 
So the middle fixed point is collinear with each of the other two points, 
but along different axes. This is precisely the reason why
the three-black hole system considered in \cite{Emparan_Reall} is not 
collinear. On the other hand, consider the three alternate fixed points
$a_1$, $a_3$ and $a_5$. The points $a_1$ and $a_3$ are joined up along
the $z$-axis by a finite $\psi$-axis and a finite $\varphi$-axis 
(corresponding to two orthogonal $S^2$'s); a similar situation occurs
between $a_3$ and $a_5$. Suppose we measure the distances 
between these three points by coordinate displacements along the two 
angular axes, with $a_{ij}\equiv|a_i-a_j|$. Then we shall refer to the 
three points as being `collinear' if the ratios $\frac{a_{12}}{a_{23}}$
and $\frac{a_{34}}{a_{45}}$ are equal. In other words, one has to cover 
the same ratio of distances along the two orthogonal directions defined 
by the $\varphi$- and $\psi$-axes, in moving between collinear points.

This notion of collinearity is naturally compatible with the 
$U(1)\times U(1)$ generalized Weyl symmetry of the spacetime.
It turns out there is another notion of collinearity in five dimensions
that was alluded to in \cite{Emparan_Reall,Elvang}: if a spacetime
possesses an $SO(3)$ spatial isometry, then points on the symmetry axis
can be regarded as collinear. (Both these notions actually coincide 
in four dimensions, since $SO(D-2)\cong U(1)^{D-3}$ when $D=4$.) 
However, since spacetimes in the generalized Weyl class will not 
possess $SO(3)$ symmetry in general, the latter notion of collinearity 
cannot be applied here. It should also be pointed out that our proposed
notion of collinearity may not be the only possible one compatible with 
$U(1)\times U(1)$ symmetry, but it is certainly one of the simplest. 
Furthermore, as we shall see below, it passes a certain consistency 
check.\footnote{See the discussion surrounding Eq.~(\ref{l}).}

It is now a straightforward matter to reintroduce the black holes into
the background spacetime, and extend our notion of collinearity to them.
Stretching between any two adjacent black holes are now two orthogonal
topological disks, which each $D^2$ terminating on a black hole event 
horizon. Otherwise, the picture is similar to that above. For definiteness, 
consider the first three black holes from the left in Fig.~2. Now, $a_{23}$ 
is the coordinate distance from the horizon of the first black hole to the 
fixed point $a_3$ along the $\psi$-axis, while $a_{34}$ is the coordinate 
distance from $a_3$ to the horizon of the second black hole along the 
$\varphi$-axis. If the ratio $\frac{a_{23}}{a_{34}}$ is the same as the 
corresponding one between the second and third black holes, namely 
$\frac{a_{56}}{a_{67}}$, then we shall refer to these three black holes 
as being collinear. Note that we are defining collinearity of the black 
holes with respect to the positions of their event horizons, rather than 
their centers, for simplicity. If collinearity is to be defined with 
respect to the centers, then one would need to take into account the 
masses (and hence radii) of the black holes.

In the interest of generality, we shall continue to keep the 
parameters of our solution $a_i$ arbitrary for the most part of this
paper. If a collinear array of black holes is desired, then the 
parameters can be specifically chosen to satisfy the conditions 
described above.

\newsection{The two-black hole solution}

Having obtained a 5D analog of the Israel-Khan solution, the next 
step is to study some of its properties in detail. We begin by focusing 
on the two-black hole case, for simplicity and also because many of the 
characteristic properties of the general solution would already be 
present in this case. Setting $N=2$ in (\ref{general metric}) yields
the metric:
\ba
\label{2 black hole metric}
{\rm d}s^2&=&-\frac{(R_1-\zeta_1)(R_4-\zeta_4)}
{(R_2-\zeta_2)(R_5-\zeta_5)}\,{\rm d}t^2
+\frac{(R_1+\zeta_1)(R_3-\zeta_3)}{R_4-\zeta_4}\,{\rm d}\varphi^2
+\frac{(R_2-\zeta_2)(R_5-\zeta_5)}{R_3-\zeta_3}\,{\rm d}\psi^2 
\cr
&&+{\rm e}^{2\gamma_0}  \frac{\sqrt{(R_2-\zeta_2)(R_5-\zeta_5)
Y_{15}Y_{45}Y_{13}Y_{34}Y_{23}Y_{35}Y_{24}Y_{12}}}
{\sqrt{(R_1-\zeta_1)(R_4-\zeta_4)}R_1R_2R_3R_4R_5Y_{14}Y_{25}}  
({\rm d}r^2+{\rm d}z^2)\,,
\ea
which has the rod structure and parameters as in Fig.~4.

\begin{figure}
\begin{center}
\includegraphics{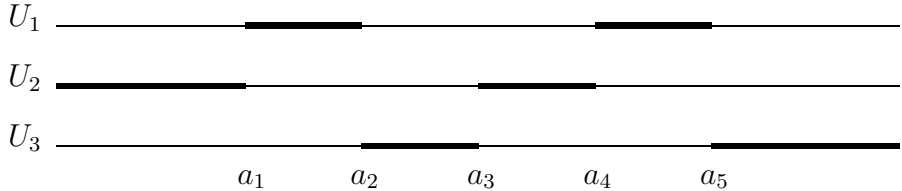}
\caption{Rod structure of the two-black hole solution.}
\end{center}
\end{figure}

Let us first check regularity conditions for the spatial sections of 
(\ref{2 black hole metric}). Consider the ${\rm d}s_{\varphi r}^2$ part of 
the metric. It turns out that conical singularities cannot be avoided along
the $\varphi$-axis, and must at least be present either along the `inner' 
part $a_3<z<a_4$ or the `outer' part $z<a_1$. By choosing 
${\rm e}^{2\gamma_0}=\frac{1}{8}$ and the period of $\varphi$ to be $2\pi$, 
we have a regular outer axis and a conical singularity running along the 
inner one. A similar situation applies to the ${\rm d}s_{\psi r}^{2}$ part 
of the metric. Explicitly, we find the conical excesses:
\alpheqn
\ba
\delta_\varphi &=& 2\pi \Bigg( \frac{a_{14}a_{25}}
{\sqrt{a_{15}a_{34}a_{35}a_{24}}}-1 \Bigg)\,, \qquad \textrm{for}~a_3<z<a_4\,;\\
\delta_\psi &=& 2\pi \Bigg( \frac{a_{14}a_{25}}
{\sqrt{a_{15}a_{13}a_{23}a_{24}}}-1 \Bigg)\,, \qquad \textrm{for}~a_2<z<a_3\,,
\ea
\reseteqn
where $a_{ij}\equiv|a_i-a_j|$ denotes the coordinate distance between
$a_i$ and $a_j$ along the $z$-axis. That conical excesses, or struts, have 
appeared between the black holes agrees with our physical intuition: they 
provide the pressure necessary to counter-balance the gravitational 
attraction of the black holes and achieve a static configuration. This 
is analogous to the 4D case \cite{Costa_Perry}, but with one important 
difference: the struts are now extended in two spatial dimensions, and 
are therefore membranes. They have the topology of disks, as described 
in Sec.~2, with their boundary circles wrapping around the black hole
event horizons.

It was also pointed out in Sec.~2 that conical singularities remain
in the background spacetime even when the black holes are removed with the
choice $a_1=a_2$ and $a_4=a_5$. In this case, the result is just the Euclidean 
C-metric solution with an added flat direction. The conical excesses 
along the inner axes are now
\alpheqn
\ba
\label{delta_phi}
\delta_\varphi &=& 2\pi\, \frac{a_{23}}{a_{34}}\,, \qquad 
\textrm{for}~ a_3<z<a_4\,;\\
\label{delta_psi}
\delta_\psi &=& 2\pi\, \frac{a_{34}}{a_{23}}\,, \qquad 
\textrm{for}~ a_2<z<a_3\,.
\ea
\reseteqn

We proceed to show in detail that our solution really consists of a 
superposition of two Schwarzschild black holes. Suppose we center
ourselves on one black hole, say the one on the left, and push the other 
infinitely far away. Note that there is an ambiguity in this procedure, 
since there are two possible directions in which this black hole can be 
pushed. We shall therefore demand that it be pushed to infinity in such 
a way that {\it it remains collinear with the original system\/}. 
In view of our notion of collinearity defined in Sec.~2, this means we
should take the limit $a_3\rightarrow\infty$ while preserving the ratio
\be
\label{l}
l\equiv\frac{a_{34}}{a_{23}}\,.
\ee
After taking this limit and performing the coordinate transformation
\alpheqn
\ba
\label{transformation1}
r&=&\frac{1}{2}\sqrt{1-\frac{2a_{12}}{(1+l)R^2}}\,
(1+l)R^2\sin2\theta\,,\\
z&=&-\frac{1}{2}\bigg(1-\frac{a_{12}}{(1+l)R^2}\bigg)
(1+l)R^2\cos2\theta\,,
\ea
\reseteqn
we recover the metric
\ba
{\rm d}s^2 &=& -\bigg( 1- \frac{2a_{12}}{ (1+l)R^2} \bigg) {\rm d}t^2 
+ \bigg( 1- \frac{2a_{12}}{ (1+l)R^2} \bigg)^{-1} {\rm d}R^2\cr
\noalign{\vskip5pt}
&&+R^2{\rm d}\theta^2 + R^2\sin^2\theta \, {\rm d} \varphi^2 
+ (1+l)^2R^2\cos^2\theta \,{\rm d}\psi^2.
\ea
This is just the 5D Schwarzschild black hole, but there is a conical 
singularity, with excess angle $2\pi l$, attached to it and stretching 
to infinity along the $\psi$-axis. It can be seen that this conical 
singularity is an artifact of the background spacetime, since the latter 
{\it has\/} a conical singularity with exactly the same excess angle 
given by (\ref{delta_psi}). This is a good consistency check, and it
shows that we have taken the infinite-distance limit correctly. Now, 
the presence of the conical singularity will affect the calculation 
of the ADM mass of this black hole \cite{Ford}, since spacetime is no 
longer asymptotically flat. Following the procedure of \cite{Ford}, 
we calculate its mass to be $\frac{3}{4}\pi a_{12}$.

In a similar fashion, we can center ourselves on the right black hole,
and push the left one to infinity. In doing so, we recover the 
limiting metric
\ba
{\rm d}s^2 &=& -\bigg( 1- \frac{2la_{45}}{ (1+l)R^2} \bigg)
{\rm d}t^2 
+ \bigg( 1- \frac{2la_{45}}{ (1+l)R^2} \bigg)^{-1} {\rm d}R^2\cr
\noalign{\vskip5pt}
&&+R^2{\rm d}\theta^2 + \bigg( \frac{1+l}{l} \bigg)^2
R^2 \sin^2\theta \, {\rm d} \varphi^2 + R^2\cos^2\theta\,{\rm d}\psi^2.
\ea
Again we obtain a Schwarzschild black hole, with a conical 
singularity now stretching to infinity along the $\varphi$-axis. 
It has excess angle $2\pi l^{-1}$, in agreement with (\ref{delta_phi}).
The mass of this black hole can be calculated to be $\frac{3}{4}\pi a_{45}$.
The sum of the masses of the two individual black holes is therefore 
\be
M=\frac{3}{4}\pi(a_{12}+a_{45})\,, 
\ee
and it turns out to be equal to the calculated ADM mass of the full 
solution (\ref{2 black hole metric}). This is to be expected since
the interaction energy between the black holes (determined by the
conical singularities \cite{Costa_Perry}) vanishes in the infinite
separation limit, and so the total energy of the system is just the 
sum of the masses of the separate black holes.

We shall now show that our solution describes two Schwarzschild 
black holes even if the distance between them is kept finite. This
involves taking the near-horizon limit of each black hole. Let us 
focus on the left black hole. If we perform the coordinate transformation
\alpheqn
\ba
\label{transformation2}
r&=&\frac{1}{2}\sqrt{1-\frac{2a_{12}}{R^2}}\,R^2\sin2\theta\,,\\
z&=&-\frac{1}{2}\bigg(1-\frac{a_{12}}{R^2}\bigg)R^2\cos2\theta\,,
\ea
\reseteqn
and then expand (\ref{2 black hole metric}) near $R=\sqrt{2a_{12}}$, 
we obtain
\ba
\label{near horizon metric}
{\rm d}s^2&=&f_1^2(\theta) \Bigg[ -\bigg(1-\frac{2a_{12}}{R^2} \bigg)
\,{\rm d}t^2
+\frac{a_{15}a_{13}}{a_{14}^2}\Bigg\{\bigg( 1-\frac{2a_{12}}{R^2} \bigg)^{-1} 
{\rm d}R^2+R^2{\rm d}\theta^2\Bigg\} \Bigg]\cr
&& + f_2^2(\theta)R^2\sin^2\theta\,{\rm d}\varphi^2 
+f_3^2(\theta)R^2\cos^2\theta\,{\rm d}\psi^2,
\ea
where
\alpheqn
\ba
f_1^2(\theta)&\equiv&\frac{a_{12}\cos^2\theta+a_{24}}{a_{12}\cos^2\theta+a_{25}}\,,
\\
f_2^2(\theta)&\equiv&\frac{a_{12}\cos^2\theta+a_{23}}{a_{12}\cos^2\theta+a_{24}}\,,
\\
f_3^2(\theta)&\equiv&\frac{a_{12}\cos^2\theta+a_{25}}{a_{12}\cos^2\theta+a_{23}}\,.
\ea
\reseteqn
The metric (\ref{near horizon metric}) describes the near-horizon geometry of 
a Schwarzschild black hole, albeit distorted away from spherical symmetry.
This angular distortion is encoded by the three so-called distortion factors 
$f_1$, $f_2$ and $f_3$, and can be attributed to the gravitational pull 
of the other black hole. It is only when the latter is pushed to infinity
do the distortion factors disappear. 

One can similarly analyze the right black hole, and would find its 
near-horizon geometry to be given by the above expressions upon switching
$\varphi\leftrightarrow\psi$, $\theta\rightarrow\frac{\pi}{2}-\theta$, 
$a_{12}\rightarrow a_{45}$, $a_{23}\rightarrow a_{34}$, etc. For simplicity, 
let us assume here that $a_{12}=a_{45}$ and $a_{23}=a_{34}$, corresponding
to a left-right symmetric system. We shall take a 
$t=R=\varphi=\psi={\rm constant}$ slice of the near-horizon metric
to see how the quarter-circle $0\leq\theta\leq\frac{\pi}{2}$, with proper 
radius $\sqrt{g_{\theta\theta}}$, of each black hole is affected by the 
other. For the left black hole, we have
\be
g_{\theta\theta} \propto \frac{\cos^2\theta+w}{1+\cos^2\theta+w} \,,
\ee
where $w\equiv\frac{a_{24}}{a_{12}}$ is a parameter related to the 
coordinate separation-to-mass ratio of the two-black hole system. It is 
readily seen (see Fig.~5) that when $w$ is large, we approach 
perfect quarter-circles for both black holes. As $w$ decreases, the two 
quarter-circles start to deviate from circular symmetry. In the limit 
$w \rightarrow 0$, the quarter-circles pinch off along the axes joining
them. This behavior is rather similar to that in the Israel-Khan 
solution \cite{Costa_Perry}.

\begin{figure}
\begin{center}
\includegraphics{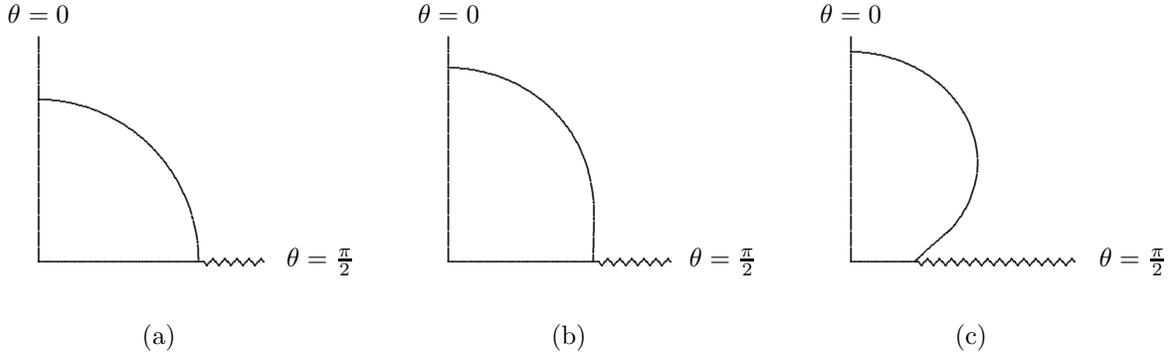}
\caption{The horizon of the left black hole as represented by the 
quarter-circle $0\leq\theta\leq\frac{\pi}{2}$, with proper radius 
$\sqrt{g_{\theta\theta}}$, for (a) $w=5$, (b) $w=0.5$, and (c) $w=0.05$. 
The jagged lines denote the conical singularities stretching between 
the two black holes.}
\end{center}
\end{figure}

Let us now turn briefly to some other properties of our two-black hole
solution. From the near-horizon metric of each black hole, one can 
calculate the 3-areas of the event horizons to be
\alpheqn
\ba
{\cal A}_{\rm{left}}&=&\sqrt{2}(2\pi)^{2}\frac{a_{12}\sqrt{a_{12}a_{13}a_{15}}}
{a_{14}}\,,\\
{\cal A}_{\rm{right}}&=&\sqrt{2}(2\pi)^{2} \frac{a_{45}\sqrt{a_{45}a_{35}a_{15}}}
{a_{25}}\,.
\ea
\reseteqn
To calculate the Hawking temperature associated with each event horizon, 
it is convenient to Euclideanize our solution $t\rightarrow -i\tau$. 
The natural period of $\tau$ is then the inverse Hawking temperature. 
We obtain
\alpheqn
\ba
T_{\rm{left}}&=&\frac{1}{\sqrt{2}2 \pi}
\frac{a_{14}}{\sqrt{a_{15}a_{13}a_{12}}}\,,\\
T_{\rm{right}}&=&\frac{1}{\sqrt{2}2 \pi}
\frac{a_{25}}{\sqrt{a_{15}a_{45}a_{35}}}\,.
\ea
\reseteqn

A question then arises if the two black holes can be in 
thermodynamic equilibrium for some choice of parameters. In the 4D case, 
it was shown in \cite{Costa_Perry} that the two black holes have to be of 
the same mass for the system to be in thermal equilibrium. To examine our 
solution likewise, we equate the expressions for the two temperatures
to find
\be
\label{quadratic}
(a_{14}^2-a_{13}a_{12})a_{45}^2+(a_{14}^2a_{34}-2a_{13}a_{12}a_{24})
a_{45}-a_{13}a_{12}a_{24}^2=0\,.
\ee
It can be shown that (\ref{quadratic}) has positive solutions for $a_{45}$ 
for any $a_{12}$, $a_{23}$ and $a_{34}$. In particular, we have the 
solution $a_{12}=a_{45}$ and $a_{23}=a_{34}$. Thus, we conclude that in 
our solution, the two black holes need not be of the same mass for them 
to have the same temperature. If they do, then the two finite rotational 
axes between them must be of the same length.

Finally we observe a Smarr relation for either black hole of our solution, 
consistent with that in \cite{Myers_Perry}:
\be
M_j=\frac{3}{2}T_{j}\bigg(\frac{{\cal A}_{j}}{4}\bigg) \,,
\ee
where $M_j$ is the mass of the black hole.
In its differential form, the Smarr relation may be identified with
the first law of black hole thermodynamics, with $\frac{{\cal A}_{j}}{4}$
the entropy of the black hole.

\newsection{The three-black hole solution}

The techniques used in the preceding section to analyze the $N=2$ case
of (\ref{general metric}) can be straightforwardly extended to any other
$N>2$. However, the various calculations will get much more tedious. In 
this section, we shall briefly study the $N=3$ solution, concentrating
on the central black hole in this system as it would exhibit some features 
not present in the $N=2$ case. 

For a three-black hole system, the metric (\ref{general metric}) reduces to
\ba
\label{3 black hole metric}
{\rm d}s^2 &=& -\frac{(R_1-\zeta_1)(R_4-\zeta_4)(R_7-\zeta_7)}
{(R_2-\zeta_2)(R_5-\zeta_5)(R_8-\zeta_8)} {\rm d}t^2 
+ (R_1+\zeta_1)\frac{(R_3-\zeta_3)(R_6-\zeta_6)}
{(R_4-\zeta_4)(R_7-\zeta_7)}{\rm d}\varphi^2 
\cr
\cr
&&+ \frac{(R_2-\zeta_2)(R_5-\zeta_5)}
{(R_3-\zeta_3)(R_6-\zeta_6)} (R_8-\zeta_8){\rm d}\psi^2
+\frac{1}{16\sqrt{2}}\frac{\sqrt{(R_2-\zeta_2)(R_5-\zeta_5)(R_8-\zeta_8)}}
{\sqrt{(R_1-\zeta_1)(R_4-\zeta_4)(R_7-\zeta_7)}}
\cr
\cr
&&\times\frac{\sqrt{Y_{37}Y_{46}Y_{26}Y_{35}Y_{45}Y_{27}Y_{18}Y_{48}Y_{38}
Y_{13}Y_{12}Y_{34}Y_{23}Y_{24}Y_{78}Y_{68}Y_{16}Y_{15}Y_{67}Y_{56}Y_{57}}}
{R_1R_2\cdots R_8 Y_{28}Y_{14}Y_{58}Y_{17}Y_{47}Y_{36}Y_{25} }
\,({\rm d}r^2 + {\rm d}z^2)\,, 
\cr&&
\ea
where the rod parameters are defined in Fig.~6, and the factor 
$\frac{1}{16\sqrt{2}}$ has been chosen to make the outer rotational axes 
$z<a_1$ and $z>a_8$ regular. As usual, there are conical excesses 
resulting along the finite inner axes, but these cannot be removed by 
any choice of parameters. They are necessary to hold the system in 
static equilibrium.

\begin{figure}
\begin{center}
\includegraphics{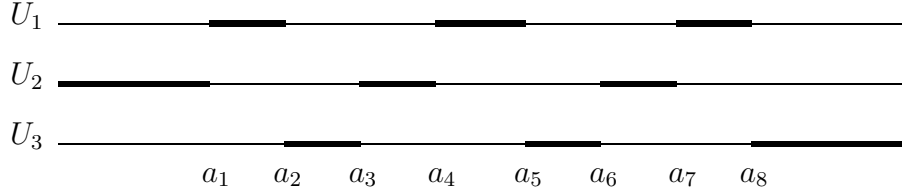}
\caption{Rod structure of the three-black hole solution.}
\end{center}
\end{figure}

We can show that (\ref{3 black hole metric}) indeed describes a 
three-black hole configuration by performing the same limiting procedures 
as in the two-black hole case. In particular, to recover the central 
black hole, we center our coordinates on it and push the other two
black holes to infinity in a collinear fashion. This is done by taking 
the limit $a_6\rightarrow\infty$ such that the three ratios
\be
l_1\equiv\frac{a_{56}}{a_{67}}\,, \qquad 
l_2\equiv\frac{a_{34}}{a_{67}}\,, \qquad 
l_3\equiv\frac{a_{24}}{a_{67}}\,,
\ee
remain fixed. After a coordinate transformation analogous to 
(3.5), the metric becomes
\ba
{\rm d}s^2 &=& -\bigg( 1- \frac{2a_{45}}{\lambda_{1} \lambda_{2} R^2}
 \bigg) {\rm d}t^2 
+ \bigg( 1- \frac{2a_{45}}{\lambda_{1}\lambda_{2} R^2} \bigg)^{-1} 
{\rm d}R^2\cr
\noalign{\vskip5pt}
&&+R^2{\rm d}\theta^2 + \lambda^2_{1} R^2\sin^2\theta\,{\rm d}\varphi^2 
+\lambda^2_{2} R^2 \cos^2\theta\, {\rm d} \psi^2,
\ea
where 
\be
\lambda_{1} \equiv\frac{l_3(l_1+l_2)(1+l_1+l_3)}{l_2(1+l_1+l_2)(l_1+l_3)}\,,
\qquad \lambda_{2}\equiv
\frac{(1+l_1)(l_1+l_2)(1+l_1+l_3)}{l_1(1+l_1+l_2)(l_1+l_3)}\,. 
\ee
Thus we recover the Schwarzschild black hole, but with conical 
singularities attached to it along two different directions. The calculated
values of the conical excesses coincide with those of the corresponding
Euclidean multiple C-metric background. 

To study the central black hole more carefully, let us consider its 
near-horizon geometry. Similar to the two-black hole case, we center 
ourselves on it and perform a coordinate transformation analogous 
to (3.9). After expanding the metric near $R=\sqrt{2a_{45}}$, we get 
\ba
\label{near horizon metric 3}
{\rm d}s^2 &=& g_1^2(\theta) \Bigg[ -\bigg(1-\frac{2a_{45}}{R^2} \bigg){\rm d}t^2 
+ B\,\Bigg\{ \bigg( 1-\frac{2a_{45}}{R^2} \bigg)^{-1} {\rm d}R^2+
R^2{\rm d}\theta^2\Bigg\} \Bigg]\cr
\noalign{\vskip5pt}
&&+ g_2^2(\theta) R^2 \sin^2 \theta\, {\rm d}\varphi^2 
+ g_3^2(\theta) R^2 \cos^2 \theta\, {\rm d}\psi^2,
\ea
where 
\be
B\equiv\frac{a_{18}a_{48}a_{38}a_{16}a_{15}a_{37}a_{46}a_{26}a_{35}a_{27}}
{(a_{28}a_{17}a_{47}a_{36}a_{25})^2}
\ee
is a rather complicated constant term. More interestingly, the angular 
distortion factors which describe how the central black hole is affected 
by the other two black holes are
\alpheqn
\ba
g_1^2(\theta) &\equiv& \frac{(a_{45}\sin^2\theta + a_{24})
(a_{45}\cos^2\theta + a_{57})}{(a_{45}\sin^2\theta + a_{14})
(a_{45}\cos^2\theta + a_{58})}\,,\\
\cr
g_2^2(\theta) &\equiv& \frac{(a_{45}\sin^2\theta + a_{14})
(a_{45}\cos^2\theta + a_{56})}{(a_{45}\sin^2\theta + a_{34})
(a_{45}\cos^2\theta + a_{57})}\,,\\
\cr
g_3^2(\theta) &\equiv& \frac{(a_{45}\sin^2\theta + a_{34})
(a_{45}\cos^2\theta + a_{58})}{(a_{45}\sin^2\theta + a_{24})
(a_{45}\cos^2\theta + a_{56})}\,.
\ea
\reseteqn

For simplicity, we now assume that $a_{12}=a_{78}$, $a_{23}=a_{67}$
and $a_{34}=a_{56}$, corresponding to a left-right symmetric system. Again, 
we shall take a $t=R=\varphi=\psi=\textrm{constant}$ slice of the metric 
(\ref{near horizon metric 3}) and observe how the quarter-circle
$0\leq\theta\leq\frac{\pi}{2}$, with proper radius 
$\sqrt{g_{\theta\theta}}$, is affected by the other two black holes. We have
\be
g_{\theta\theta} \propto \frac{(\sin^2\theta + w)(\cos^2\theta + w)}
{(\sin^2\theta + w + v)(\cos^2\theta + w+v)}\,,
\ee
where $w\equiv\frac{a_{24}}{a_{45}}$ and $v\equiv\frac{a_{12}}{a_{45}}$.
This function encodes the distortion of the central black hole's horizon 
along the $\theta$-direction by the other two black holes. We can see
from it, even without the aid of graphical plots, the characteristic 
effects of the various physical parameters as follows: Firstly, note 
that $w$ represents the ratio of the coordinate distance between the left 
and central black holes to the mass of the latter. As it increases, the 
quarter-circle tends more towards a circular arc. Secondly, $v$ represents
the ratio of the mass of the left black hole to that of the central one.
As it increases, the quarter-circle deviates more from circular symmetry. 
This general behavior conforms to our Newtonian expectations.

Finally, we briefly study the temperatures of the black holes in this 
solution. They are
\alpheqn
\ba
T_{\rm{left}}&=&\frac{1}{\sqrt{2}2\pi}
\frac{a_{14}a_{17}}{\sqrt{a_{18}a_{13}a_{12}a_{15}a_{16}}}\,,
\\
T_{\rm{right}}&=&\frac{1}{\sqrt{2}2\pi}
\frac{a_{28}a_{58}}{\sqrt{a_{18}a_{48}a_{38}a_{78}a_{68}}}\,,
\\
T_{\rm{central}}&=&\frac{1}{\sqrt{2}2\pi}\frac{a_{28}a_{17}a_{47}a_{36}a_{25}}
{\sqrt{a_{18}a_{48}a_{38}a_{16}a_{15}a_{37}a_{46}a_{26}a_{35}a_{45}a_{27}}}\,.
\ea
\reseteqn
The question then arises if the three black 
holes can be in thermodynamic equilibrium for some choice of parameters. 
Indeed, there exists infinitely many solutions for such a scenario, a 
particular solution being 
$a_{67}=a_{56}=a_{34}=a_{23}=\frac{1}{2}a_{78}=\frac{1}{2}a_{12}$,
and $a_{45} \simeq 1.5886\,a_{67}$. In this case, the central black
hole has to have a smaller mass than the other two black holes, in 
order to achieve thermodynamic equilibrium.

\newsection{Multiple charged black holes}

We shall now generalize our solution (\ref{general metric}) to a system 
of multiple charged black holes in a general 5D Einstein-Maxwell-dilaton 
theory. This would be a first step towards embedding the solution in a
more complete framework such as string or M-theory, which may then
provide some insights into the microscopic description of such a system.

We begin by finding the corresponding multi-black hole solution in the 
special case of 5D Kaluza-Klein theory. This can be done using the standard 
procedure \cite{Frolov} of embedding the spacetime (\ref{general metric}) 
in six dimensions by adding a flat extra dimension:
\be
{\rm d}s_{(6)}^2 = -{\rm e}^{2U_1}{\rm d}t^2 
+ {\rm e}^{2U_2}{\rm d}\varphi^{2} +{\rm e}^{2U_3}{\rm d}\psi^2 
+{\rm e}^{2\nu}({\rm d}r^2 + {\rm d}z^2) + {\rm d}y^2.
\ee
Boosting along the $y$-direction with rapidity $\sigma$, the metric becomes
\ba
\label{boosted}
{\rm d}s_{(6)}^2 &=& -\frac{{\rm e}^{2U_1}}{\cosh^2\sigma 
- {\rm e}^{2U_1}\sinh^2\sigma}{\rm d}t^2 + {\rm e}^{2U_2}{\rm d}\varphi^{2} 
+{\rm e}^{2U_3}{\rm d}\psi^2 +{\rm e}^{2\nu}({\rm d}r^2 + {\rm d}z^2)\,
\cr
&&+(\cosh^2\sigma - {\rm e}^{2U_1}\sinh^2\sigma) \bigg( {\rm d}y 
- \frac{(1-{\rm e}^{2U_1})\sinh \sigma\cosh \sigma}
{\cosh^2\sigma - {\rm e}^{2U_1}\sinh^2 \sigma}{\rm d}t \bigg)^2.
\ea
If we dimensionally reduce on $y$ using the ansatz
\be
{\rm d}s_{(6)}^2 = {\rm e}^{-\frac{1}{\sqrt{6}}\phi} {\rm d}s_{(5)}^2 
+ {\rm e}^{\sqrt{\frac{3}{2}}\phi}({\rm d}y - 2A_{a}{\rm d}x^{a} )^2,
\ee
then the 5D metric ${\rm d}s_{(5)}^2$, Abelian gauge field $A_a$ and 
dilaton $\phi$ can be read off from (\ref{boosted}). They describe an 
electrically charged multi-black hole solution in 5D Kaluza-Klein theory 
with the action
\be
\label{action}
I=\frac{1}{16\pi}\int{\rm d}^5x\sqrt{-g}\bigg(R-\frac{1}{2}\partial_a\phi\partial^a\phi
-{\rm e}^{\sqrt{\frac{8}{3}}\phi}F_{ab}F^{ab}\bigg),
\ee
where $F_{ab}\equiv\partial_a A_b - \partial_b A_a$. 

Now (\ref{action}) belongs to a general class of Einstein-Maxwell-dilaton 
theories with the action
\be
\label{action2}
I=\frac{1}{16\pi}\int{\rm d}^5x\sqrt{-g}\bigg(R
-\frac{1}{2}\partial_a\phi\partial^a\phi
-{\rm e}^{\alpha\phi}F_{ab}F^{ab}\bigg),
\ee
where $\alpha$ is a constant parameterizing the coupling of the dilaton
to the gauge field. In particular, the Einstein-Maxwell 
case is recovered when $\alpha=0$. It is fairly straightforward to 
generalize our static multi-black hole solution in Kaluza-Klein theory to 
one of (\ref{action2}). The general-$\alpha$ solution turns out to be
\alpheqn
\ba
\label{general charged solution}
{\rm d}s^2 &=& -H^{-\frac{2\beta}{3}}{\rm e}^{2U_1}{\rm d}t^2 
+ H^{\frac{\beta}{3}} \bigg({\rm e}^{2U_2}{\rm d}\varphi^{2} 
+{\rm e}^{2U_3}{\rm d}\psi^2 +{\rm e}^{2\nu}({\rm d}r^2 + {\rm d}z^2) \bigg)\,,\\
A_{t} &=& \frac{\sqrt{\beta}}{2}H^{-1}(1-{\rm e}^{2U_1})\sinh \sigma\cosh \sigma\,,\qquad 
{\rm e}^{\phi} = H^{\frac{\beta \alpha}{2}},
\ea
\reseteqn
where 
\be
\beta \equiv \frac {12}{4+3\alpha^2}\,,\qquad
H\equiv1+ \sinh^2 \sigma\, \big(1-{\rm e}^{2U_1} \big)\,,
\ee
and $U_\alpha$, $\nu$ can be read off from (2.3) and (\ref{general metric}), 
respectively. The corresponding magnetically charged solution can be 
obtained by the usual electromagnetic duality transformation.

The rod structure of this solution is still given by Fig.~2. Its
ADM mass, electric and scalar charge \cite{Gibbons_Maeda}
can be calculated in terms of the masses, electric and scalar
charges of the individual black holes, as follows:
\ba
M_{\rm{total}} &=& \sum_{j=1}^N M_{j} = \frac{3\pi}{8}
\bigg(1+\frac{2\beta}{3} \sinh^2 \sigma \bigg) \sum_{j=1}^N \mu_j\,,\\
Q_{\rm{total}} &=& \sum_{j=1}^N Q_{j} = \pi^2 \sqrt{\beta} \sinh(2\sigma) 
\sum_{j=1}^N \mu_j\,,\\
\Sigma_{\rm{total}} &=& \sum_{j=1}^N \Sigma_{j} 
= 2\pi^2 \beta \alpha \sinh^2 \sigma \sum_{j=1}^N \mu_j\,.
\ea
Here we have labeled the $j^{\rm{th}}$ black hole from the left, and 
set $\mu_j\equiv2\,| a_{3j-2} - a_{3j-1}|$. Note that the individual
black holes all have the same mass-to-charge ratio.
In accordance with the no-hair 
theorem \cite{Bekenstein}, we observe as usual that the scalar charge is 
not an independent parameter, and it vanishes when the electric charge
does so because
\be
Q_j^2 = \Sigma_j \bigg( \frac {16\pi M_j}{3 \alpha} 
+ \frac{3-2\beta}{3 \beta \alpha^2} \Sigma_j \bigg)\,.
\ee
Furthermore we have a relation between the rod-length and the mass
and charge of the black hole in question, given by
\be
\mu_j^2=\bigg( \frac{8M_j}{3\pi} + \frac{\alpha \Sigma_j}{4\pi^2} \bigg)^2 
- \bigg( \frac{Q_j}{\pi^2 \sqrt{\beta}} \bigg)^2.
\ee
The limit of vanishing rod-lengths $\mu_j\rightarrow0$ is the so-called 
extremal limit. In view of its relative importance, this case would be 
discussed separately in Sec.~6.

Let us now specialize to the two-black hole case. If we center ourselves
on the left black hole and push the right one to infinity as in
Sec.~3, we obtain
\alpheqn
\ba
{\rm d}s^2&=&-\bigg( 1 + \frac{\tilde{\mu}_1}{R^2} \sinh^2 \sigma \bigg)^{-\frac{2\beta}{3}} \bigg( 1 - \frac{\tilde{\mu}_1}{R^2} \bigg) {\rm d}t^2 + \bigg( 1 + \frac{\tilde{\mu}_1}{R^2} \sinh^2 \sigma \bigg)^{\frac{\beta}{3}}
\Bigg[ \bigg(1 - \frac{\tilde{\mu}_1}{R^2} \bigg)^{-1} {\rm d}R^2 
\cr
&&+R^2{\rm d}\theta^2 + R^2 \sin^2\theta \, {\rm d} \varphi^2
+ (1+l)^2 R^2\cos^2\theta\,{\rm d}\psi^2  \Bigg]\,,\\
A_t&=&\frac{\sqrt{\beta}}{2}\frac{\tilde{\mu}_1\sinh\sigma\cosh\sigma}
{R^2+\tilde{\mu}_1\sinh^2\sigma}\,,\qquad
{\rm e}^{\phi}=\bigg( 1 + \frac{\tilde{\mu}_1}{R^2} \sinh^2 \sigma \bigg)^{\frac{\beta\alpha}{2}},
\ea
\reseteqn
where $\tilde{\mu}_1 \equiv \frac{\mu_1}{1+l}$. Thus, we recover in this 
limit a single dilatonic black hole \cite{Gibbons_Maeda}, except for a 
conical singularity attached to it along the $\psi$-axis with excess
angle $2\pi l$. In a procedure similar to that in Sec.~3, we may also 
calculate its near-horizon geometry to obtain
\ba
\label{near horizon metric of charged}
{\rm d}s^2&=&f_1^2(\theta) \Bigg[ -\cosh^{-\frac{4\beta}{3}} \sigma \,
\bigg(1-\frac{\mu_1}{R^2} \bigg)\,{\rm d}t^2  
+ \cosh^{\frac{2\beta}{3}}\sigma\,\frac{a_{15}a_{13}}{a_{14}^2} 
\Bigg\{\bigg( 1-\frac{\mu_1}{R^2} \bigg)^{-1} {\rm d}R^2
+R^2{\rm d}\theta^2 \Bigg\}\Bigg]
\cr
&&+\cosh^{\frac{2\beta}{3}} \sigma\, \bigg(
f_2^2(\theta)R^2\sin^2\theta\,{\rm d}\varphi^2
+f_3^2(\theta)R^2\cos^2\theta\,{\rm d}\psi^2 \bigg)\,,
\ea
where the distortion factors $f_1$, $f_2$, $f_3$ are the same as those in 
the vacuum case (3.11). The 3-area and temperature of the event horizon 
are respectively
\ba
{\cal A} &=& \sqrt{2}(2\pi)^{2}\cosh^{\beta}\sigma\,
\frac{a_{12}\sqrt{a_{12}a_{13}a_{15}}}{a_{14}}\,,
\\
T &=& \frac{\cosh^{-\beta}\sigma}{\sqrt{2}2\pi}
\frac{a_{14}}{\sqrt{a_{15}a_{13}a_{12}}}\,,
\ea
while the electrostatic potential at the horizon is
\be
\Phi_{\rm{horizon}} = \frac{\sqrt{\beta}}{2} \tanh \sigma\,.
\ee
Note that $\Phi_{\rm{horizon}}$ is independent of which black hole
we are considering.

A similar analysis can be performed on the right black hole, but we will
not reproduce the results here. We end off by remarking that
a generalized Smarr relation holds for the individual black holes:
\be
M_j = \frac{3}{2}T_j\bigg(\frac{{\cal A}_j}{4}\bigg)
+ \frac{\Phi_{\rm{horizon}}Q_j}{4\pi}\,.
\ee
This relation can be explicitly checked for the left black hole using the
above results, and is consistent with the Smarr formula for the 
electrically charged black holes found in 
\cite{Gibbons_Maeda}.\footnote{There is a $4\pi$ in the denominator 
of the second term because the Maxwell term in (\ref{action2}) is 
$\frac{1}{4\pi}$ times that in \cite{Gibbons_Maeda}.}

\newsection{Extremal black holes}

The extremal limit of the charged multi-black hole solution derived in
the preceding section, is taken by sending $\mu_j\rightarrow0$ and 
$\sigma\rightarrow\infty$ to infinity such that the charges $Q_j$ remain 
fixed. The solution (5.6) becomes in this limit 
\alpheqn 
\ba
\label{extremal charged solution}
{\rm d}s^2 &=& - \bigg( 1 + \sum_{j=1}^N \frac{\tilde{Q}_j}{2R_{2j-1}} \bigg)^{-\frac{2\beta}{3}}
{\rm d}t^2 + \bigg( 1 + \sum_{j=1}^N \frac{\tilde{Q}_j}{2R_{2j-1}} \bigg)^{\frac{\beta}{3}}
\cr
&&\times \Bigg[ (R_1+\zeta_1) \prod_{j=1}^{N-1} \bigg( \frac{R_{2j}-\zeta_{2j}}{R_{2j+1}-\zeta_{2j+1}} \bigg) 
{\rm d}\varphi^2 + (R_{2N-1} - \zeta_{2N-1}) \prod_{j=1}^{N-1} \bigg( \frac{R_{2j-1}-\zeta_{2j-1}}{R_{2j}-\zeta_{2j}} \bigg) {\rm d\psi^2} 
\cr&&+
{\rm e}^{2\gamma_0}   \prod_{k=1}^{N-1}\bigg( \frac{Y_{1,2k}}{Y_{1,2k+1}} \bigg)  \frac{ \prod_{l,m}^{N-1}Y_{2l+1,2m} }{\prod_{i=1}^{2N-1}R_i\, \prod_{j<s}^{N-1}Y_{2j,2s}Y_{2j+1,2s+1}} ({\rm d}r^2+{\rm d}z^2) \Bigg]\,,\\
A_t&=&-\frac{\sqrt{\beta}}{2}\bigg( 1 + \sum_{j=1}^N \frac{\tilde{Q}_j}{2R_{2j-1}} \bigg)^{-1},
\qquad{\rm e}^{\phi} = \bigg( 1 + \sum_{j=1}^N \frac{\tilde{Q}_j}{2R_{2j-1}} \bigg)^{\frac{\beta \alpha}{2}},
\ea
\reseteqn
where $\tilde{Q}_j \equiv \frac{Q_j}{2\pi^2 \sqrt{\beta}}$. Its 
corresponding rod structure is given by Fig.~3, but with the addition of
point sources for the time coordinate at $z=a_{2j-1}$, where the $N$ 
black holes are located.

Note that the part of the metric in square brackets is the Euclidean 4D 
multiple C-metric solution. This is in contrast to the 5D multi-extremal 
black hole solution previously considered in \cite{Duff}, in which the metric 
in the square brackets is the flat one. Our solution is more complicated due
to the fact that we are adding black holes to the non-trivial background
of Fig.~3, instead of flat space. However, we believe it is still worth 
studying the solution (6.1), since the non-extremal generalization of 
the multi-black hole solution of \cite{Duff} is not known. 

The ADM mass and scalar charge of the $j^{\rm th}$ extremal black hole are
expressed in terms of $Q_j$ by 
\ba
M_j &=& \frac{\sqrt{\beta} Q_j}{8\pi}\,,
\\
\Sigma_j &=& \sqrt{\beta} \alpha Q_j\,,
\ea
with the total mass and scalar charge of the solution given by their 
respective sums. 

Let us again consider the two-black hole case for simplicity. The metric 
(\ref{extremal charged solution}) reduces to
\ba
{\rm d}s^2 &=& - \bigg( 1 + \frac{\tilde{Q}_1}{2R_1} 
+ \frac{\tilde{Q}_2}{2R_3} \bigg)^{-\frac{2\beta}{3}}
{\rm d}t^2 + \bigg( 1 + \frac{\tilde{Q}_1}{2R_1} 
+ \frac{\tilde{Q}_2}{2R_3} \bigg)^{\frac{\beta}{3}} 
\,\Bigg[\frac {(R_1+\zeta_1)(R_2-\zeta_2)}{R_3-\zeta_3}  {\rm d}\varphi^2
\cr
&& +\frac {(R_1- \zeta_1)(R_3-\zeta_3)}{R_2-\zeta_2}  {\rm d}\psi^2 
+ \frac{Y_{12}Y_{23}}{4R_1R_2R_3Y_{13}}({\rm d}r^2 + {\rm d}z^2)\Bigg]\,,
\ea
where the part of the metric in square brackets is just the usual 
Euclidean C-metric solution \cite{Bonnor}. The black holes are located at 
$z=a_1$ and $a_3$. Centering on the left black hole and pushing 
the other to infinity as was done above, we obtain 
\ba
{\rm d}s^2&=&-\bigg( 1 + \frac{\tilde{Q}_1}{(1+l)R^2}  \bigg)^{-\frac{2\beta}{3}} {\rm d}t^2 + \bigg( 1 + \frac{\tilde{Q}_1}{(1+l)R^2} \bigg)^{\frac{\beta}{3}}
\bigg[ {\rm d}R^2 
+R^2{\rm d}\theta^2 \cr
&&+  R^2 \sin^2\theta\, {\rm d} \varphi^2+(1+l)^2R^2\cos^2\theta\,{\rm d}\psi^2 \bigg],
\ea
where now $l\equiv\frac{a_{23}}{a_{12}}$. This is the extreme dilatonic 
black hole metric of \cite{Gibbons_Maeda,Duff}, but with a conical 
singularity attached to the $\psi$-axis. Its near-horizon limit is simply 
\ba
\label{extremal near-horizon}
{\rm d}s^2&=&-\bigg(  \frac{\tilde{Q}_1}{(1+l)R^2}  \bigg)^{-\frac{2\beta}{3}} {\rm d}t^2 + \bigg(  \frac{\tilde{Q}_1}{(1+l)R^2} \bigg)^{\frac{\beta}{3}}
\bigg[ {\rm d}R^2 
+R^2{\rm d}\theta^2 \cr
&&+ R^2 \sin^2\theta\, {\rm d} \varphi^2+(1+l)^2 R^2\cos^2\theta\,{\rm d}\psi^2\bigg].
\ea
The respective limits for the right black hole are similar, with the conical 
singularity attached to the $\varphi$-axis instead. Note that there is an 
absence of angular distortion in (\ref{extremal near-horizon}). This is
due to the well-known fact that the electrostatic repulsion exactly balances 
the gravitational attraction between extremally charged black holes. There
are however, conical singularities still stretching between the black holes,
but these are intrinsic to the background spacetime and cannot be avoided.

The areas ${\cal A}_i$ of the event horizons are zero except for the 
$\beta = 3$ (Einstein-Maxwell) case where we find
\alpheqn 
\ba
{\cal A}_{\rm{left}} &=& \sqrt{\frac{1}{2\pi^2}\frac{a_{12}}{a_{13}}} \bigg( \frac{Q_1}{\sqrt{3}} \bigg)^{\frac{3}{2}}\,,\\
{\cal A}_{\rm{right}} &=& \sqrt{\frac{1}{2\pi^2}\frac{a_{23}}{a_{13}}} \bigg( \frac{Q_2}{\sqrt{3}} \bigg)^{\frac{3}{2}}\,.
\ea
\reseteqn
Furthermore, it is interesting to observe, as in \cite{Gibbons_Maeda}, 
the dependence of the Hawking temperatures $T$ of the two extremal black 
holes on the strength of the coupling constant:
\alpheqn 
\ba
&&T_{\rm{both}} \rightarrow \infty\,, \qquad {\rm for}~0< \beta <1\,;
\\ 
&&T_{\rm{both}} = 0\,,~~ \qquad {\rm for}~1 < \beta \leq 3\,;
\\
\noalign{\vskip5pt}
&&\left.\matrix{T_{\rm{left}} =\sqrt{\frac{1}{2Q_1}\frac{a_{13}}{a_{12}}} \,,\cr
T_{\rm{right}} =\sqrt{\frac{1}{2Q_2}\frac{a_{13}}{a_{23}}}\,,}\right\}\qquad {\rm for}~\beta=1\,.
\ea
\reseteqn
For $0<\beta<1$, the extremal limit brings the temperature to formal 
infinity, similar to the behavior of 4D Kaluza-Klein extremal black holes 
\cite{Gibbons_Wiltshire}. It was shown that these infinitely hot 
extremal black holes are protected by mass gaps or potential barriers 
which insulate them externally, and thus they can be treated as 
elementary particles \cite{Wilczek}. For $1<\beta \leq 3$, the 
temperature tends to zero smoothly, characteristic of extremal
Einstein-Maxwell black holes which are stable endpoints of black 
hole evaporation 
\cite{Garfinkle}. The $\beta=1$ case has a finite temperature. This 
enigmatic case emerges from low-energy effective string theory, when 
compactified to five dimensions. The finite temperature might lead one to 
think that the extremal endpoint of black hole evaporation will result in 
the formation of a naked singularity, but there exists various arguments to 
avoid this conclusion \cite{{Horo},{Giddings}}.

\newsection{Discussion}

In this paper, we have constructed a static solution describing a 
superposition of $N$ Schwarzschild black holes, which may be considered 
a 5D generalization of the Israel-Khan solution. For certain choices of 
parameters, the black holes may be regarded as collinear. The main 
properties of these solutions were then studied. While they share 
many properties with the Israel-Khan solution, there are also crucial
differences, particularly in the structure of the conical singularities.
The charged generalization of this solution was also considered.

There are a number of avenues for further research. For example, the 
interaction between two 4D near-extremal black 
holes was analyzed in \cite{Costa_Perry}, by embedding them in M-theory as 
bound states of branes. Using an effective string description of these bound 
states, the semi-classical result for the entropy, and its correction 
due to the interaction between the black holes, was reproduced for 
large separation. It would be very interesting to see if an effective
string description can also be found for our 5D charged two-black hole 
solution.

In four dimensions, there exists a class of solutions known as black
diholes \cite{Emparan}, which consist of pairs of black holes with 
equal mass, and
charges of the same magnitude but opposite sign. This is in contrast
to the multi-charged black hole solutions of \cite{Costa_Perry} and in
this paper, whose black holes all carry charges of the same sign. An
effective string model for near-extremal black diholes was found in
\cite{Emparan_Teo}, in terms of an interacting system of strings and
anti-strings. A natural question is whether these results would
generalize to five dimensions. A first step in this direction was 
recently made in \cite{Teo}, in which a 5D extremal black dihole 
solution was found using the generalized Weyl formalism. Like the
two-black hole solutions considered in this paper, the black holes 
exist in the background of the Euclidean C-metric.

We note that by removing all the finite rod sources for the $\varphi$ 
coordinate and the left-most rod source for the time coordinate in Fig.~2, 
we obtain a limiting metric describing multiple concentric black rings. 
This solution can be analyzed almost in parallel with the multi-black
hole solution of this paper. Another possible black ring
configuration that one could consider, is obtained from the two-black
hole rod structure (Fig.~4) by moving the finite rod source for the
$\varphi$ coordinate to the $\psi$ coordinate, and {\it vice versa\/}.
The resulting solution describes a pair of orthogonal black rings. 
Superpositions of black rings and black holes are also possible.

Finally, there remains the open question of whether it is possible to
construct a multi-black hole solution in five dimensions with $SO(3)$
instead of $U(1)\times U(1)$ symmetry. As mentioned in Sec.~2, such a
solution would possess one symmetry axis rather than two, and so would
in some sense resemble the Israel-Khan solution more closely. 
However, to construct such a solution requires one 
to move beyond the generalized Weyl formalism. Unfortunately, there has 
been little headway in this direction so far, mainly because the Einstein 
equations are no longer reducible to a linear equation, as in (\ref{Laplace}).

\bigbreak\bigskip\bigskip\centerline{{\bf Acknowledgement}}
\nobreak\noindent We would like to thank Roberto Emparan for his comments 
and suggestions.

\bigskip\bigskip

{\renewcommand{\Large}{\normalsize}
}
\end{document}